\documentclass[osajnl,twocolumn,showpacs,superscriptaddress,10pt]{revtex4-1} 
\usepackage{amsmath,amssymb,graphicx}
\usepackage{xcolor}
\begin{document}

\title{Improving Photoacoustic-guided Focusing in Scattering Media by Spectrally Filtered Detection}

\author{Thomas Chaigne}
\author{Ori Katz}
\author{J\'er\^ome Gateau}
\author{Claude Boccara}
\author{Sylvain Gigan}
\author{Emmanuel Bossy}\email{Corresponding author: emmanuel.bossy@espci.fr}

\affiliation{Institut Langevin, ESPCI ParisTech, CNRS UMR 7587, 1 rue Jussieu, 75005 Paris, France}

\begin{abstract}
We experimentally and numerically study the potential of photoacoustic-guiding for light focusing through scattering samples via wavefront-shaping and iterative optimization. We experimentally demonstrate that the focusing efficiency on an extended absorber can be improved by iterative optimization of the high frequency components of the broadband photoacoustic signal detected with a spherically focused transducer. We demonstrate more than 8-fold increase in the photoacoustic signal generated by a 30 $\mu m $ wire using a narrow frequency band around 60MHz. We numerically confirm that such optimization leads to a smaller optical focus than using the low frequency content of the photoacoustic feedback.
\end{abstract}

\maketitle 

Focusing light, ideally down to the micrometer scale, is required in many biomedical and industrial applications such as optical microscopy and light delivery. However, light scattering in complex media such as biological tissue prevents focusing and imaging at depth with optical resolution. As a result, optical microscopy is currently restricted to rather superficial investigation, i.e hundreds of micrometers\cite{ntziachristos2010going}. 

Nevertheless, in 2007, Vellekoop et al. have demonstrated focusing of coherent light to a diffraction-limited spot through highly scattering samples by wavefront shaping\cite{vellekoop2007focusing}. By controlling the phase of the incident wavefront with a spatial light modulator (SLM), they demonstrated control over the scattered light complex interference speckle pattern. Iteratively optimizing a feedback signal from a chosen target position resulted in a diffraction limited, high-intensity focus at this point.

Following this demonstration, wavefront-shaping has been exploited in numerous works to achieve various goals, from surpassing the diffraction limit in scattering media to allowing imaging through opaque layers\cite{mosk2012controlling}. In most of these works the feedback signal was obtained from a photodetector placed at the targeted location. 
However, to focus light inside a  scattering sample in most practical scenarios, one cannot directly place such a detector at the target position. 

One promising approach to perform controlled focusing inside scattering media was presented by Kong et al.\cite{kong2011photoacoustic}. This approach is based on the photoacoustic effect in the transient regime, i.e. the generation of broadband ultrasonic waves subsequent to the absorption of a short optical pulse. The  amplitude of the generated ultrasound waves is proportional to the local light intensity on the absorber. Furthermore, biological tissue already contain endogenous absorbers such as hemoglobin, and are only weakly scattering for ultrasound waves in the megahertz range. Hence, photoacoustics is an effective approach for localizing absorbing structures inside tissue, and monitoring the absorbed light intensity by detecting the emitted ultrasound waves with an external transducer\cite{ntziachristos2005looking}.
By optimizing such a photoacoustic feedback signal, it is possible to focus light through a scattering layer noninvasively , as demonstrated by Kong et al.\cite{kong2011photoacoustic}. 

Inside tissue, one will likely deal with extended absorbers (larger than the acoustic focus in at least one dimension), such as blood vessels\cite{zhang2009vivo}. As a consequence, the size of the optical focus after optimization is linked to the acoustic resolution, and the optical intensity enhancement over the probed volume is expected to drop in proportion to the number of speckle grains contained in the acoustic focus, and whose intensities are simultaneously enhanced\cite{vellekoop2008demixing,2013arXiv1305.6246C}.
In this letter, we investigate the role of the center frequency and bandwidth of the ultrasound detection on  the optical focusing capabilities (i.e. focus size and enhancement of the optical intensity on the target), considering the use of a single high frequency focused ultrasonic transducer and an extended absorber. 

We experimentally demonstrate a significant improvement in the photoacoustic signal enhancement by optimizing the high frequency components of the photoacoustic signal rather than  the full band. We also confirm with numerical simulations that such spectral filtering leads to tighter optical foci compared to the ones obtained with optimization of the low frequency signal. \newline

\begin{figure}[h!]
\centerline{\includegraphics[width=1\columnwidth]{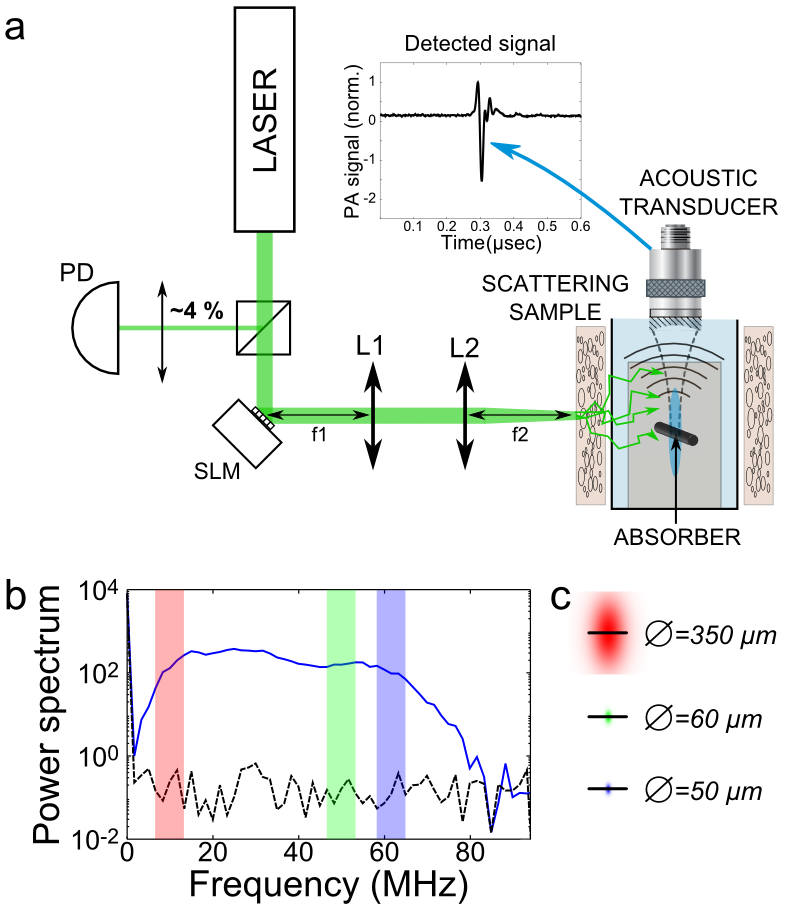}}
\caption{(a) Experimental setup: a SLM shapes a nanosecond pulsed laser beam illuminating an absorbing wire through a scattering sample; a small portion of the beam is directed toward a photodiode (PD)  to monitor the pulse to pulse intensity variation; a spherically focused acoustic transducer detects the photoacoustic signals generated by the absorber. The focal zone of the transducer is depicted by a blue elliptic region. (inset: a typical photoacoustic signal trace by the transducer). (b) Measured photoacoustic power spectrum (blue curve), and noise-floor of the detection system (dashed-black curve); the colored strips correspond to the frequency bands selected for optimization. (c) Typical dimensions of the acoustic foci corresponding to each frequency band: 5-12MHz (red), 45-52MHz (green) and 58-65MHz (blue). The black line depicts the absorbing wire.}
\label{montage}
\end{figure}

For our experimental investigation, we have implemented the following iterative optimization process to focus light using photoacoustic feedback. In this process, we sequentially shift the phase of a set of wavefront patterns that form a basis of the SLM pixels (i.e. the optical input modes). We record the modulation of the photoacoustic signal from the chosen position, and then keep the phase that maximizes it for each input mode(see Fig.\ref{montage}).
The optimization algorithm is using Hadamard basis vectors as the input basis instead of the canonical one (i.e. pixel-by-pixel), which allows for a larger modulation of the feedback signal and fast convergence\cite{popoff2010measuring}.

The experimental setup was presented in Fig.1a: a 5ns pulsed laser beam (Continuum Surelite, 10 Hz repetition rate, 532nm wavelength, $\leqslant 10mJ$ pulse energy, a few millimeters of coherence length), was expanded to illuminate a phase-only SLM (Boston Micromachines Multi-DM with 140 segmented mirrors).  
The scattering medium was an optical diffuser ($0.5^\circ$ circular light shaping diffuser, Newport). The SLM surface was imaged on the surface of the diffuser by a 4-f telescope. This ensured that the optical speckle size on the target remained constant for different SLM phase-patterns during the optimization\cite{vellekoop2010exploiting}. 
The absorbing target was a black $30\mu m$ diameter nylon wire (NYL02DS, 10/0, Vetsuture), embedded in an agarose gel placed at a distance of $5cm$ behind the scattering sample. The beam diameter on the diffuser and the distance between the diffuser and the absorber were set to obtain a speckle grain size of $25\mu m$. 
The photoacoustic signal generated by the absorber was detected with a spherically-focused ultrasonic transducer (Sonaxis) having a center frequency $f=27 MHz$ and $B=26 MHz$ two-way -6dB bandwidth. 
The detector has an intrinsic focal zone with a transverse diameter of $\varnothing_{acoustic} \simeq \textstyle \frac{F c}{f D}=100\mu m$, where $F=8mm$ is the focal length of the transducer, $D=4mm$ its diameter, $c_{\mathit{water}}=1480 m/s$ the speed of sound in water. It is worth noting that this transverse diameter is inversely proportional to the acoustic center frequency\cite{kino1987acoustic}, and can therefore be controlled by approprate filtering. 
The absorber was placed perpendicular both to the laser beam propagation direction and the transducer axis, one portion being in the acoustic focus (see Fig.\ref{montage}a). For each laser pulse and displayed SLM phase-pattern, the  photoacoustic signal was measured by a computer-controlled oscilloscope (Lecroy Wavesurfer 104MXs-B), giving the temporal photoacoustic trace (Fig.\ref{montage}, inset).
The Fourier transform of the measured signal was computed and the RMS value in a selected bandwidth was used as the feedback signal for the iterative optimization algorithm. The feedback signal was normalized by the power of each pulse (as monitored by a photodiode) to compensate for the fluctuations of the laser. \newline

To quantify the efficiency of the optimization, we define the enhancement $\eta$ of the feedback signal as the ratio between the final optimized value of the feedback signal and the mean value of the feedback signal over the first iteration of the optimization. 
In previous works using direct detection of the light intensity at the target point\cite{vellekoop2008demixing,popoff2011controlling}, the enhancement $\eta_{\mathit{optical}}$ was given by:
\begin{equation}
\eta_{\mathit{optical}} \simeq 0.5\times \dfrac{N_{\mathit{SLM}}}{N_{\mathit{modes}}} \ \ ,
\label{etaopt}
\end{equation}
where $N_{\mathit{SLM}}$  is the number of controllable degrees of freedom, and $N_{\mathit{modes}}$ is the number of speckle grains or optical modes whose intensities are simultaneously optimized.

For the photoacoustic feedback optimization considered here, we expect to obtain a similar dependance on the number of simultaneously optimized speckle grains. These speckle grains correspond to the ones contained in the so-called acoustic cell\cite{2013arXiv1308.0243G}, i.e. the absorber portion that is contained in the focal volume of the transducer. Because these speckle grains are assumed to be invariant in the propagation direction, this number scales as: $N_{\mathit{modes}} \propto \varnothing_{acoustic} \times \varnothing_{absorber} \propto f^{-1}$, in the specific case of a linear absorber.
To obtain larger signal enhancement, one has either to increase the number of degrees of control ($N_{\mathit{SLM}}$), which would require longer acquisition times, or reduce the number of optical modes inside the probed acoustic volume $N_{\mathit{modes}}$.
As the bandwidths of the transducer and photoacoustic signal are broad (spanning more than two octaves, see Fig.\ref{montage}b), one could probe significantly smaller focal areas by highpass filtering the detected signals.  Optimizing high spectral component of the photoacoustic response of the absorber thus allows to effectively reduce the detection volume, which would result in a smaller focal spot and an increased light delivery efficiency.
  
In addition to the above simple considerations, we emphasize that specific characteristics of photoacoustic detection (compared to a direct optical detection) may affect the optimization process.  First of all, the detected signal results from the coherent summation of the acoustic waves emitted by all absorbed speckle grain.
Then, its frequency content relies on the source geometry, that is the spatial distribution of the illumination pattern.
Moreover, the lateral detection profile of the transducer gives different weights to the speckle grains. This feature has been used in a recent paper to reach sub-acoustic resolution focusing after a genetic optimization process\cite{2013arXiv1310.5736C}, but it is beyond the scope of our study.

Despite these subtleties of photoacoustics, we still expect the photoacoustic feedback optimization to exhibit a similar trend concerning $N_{\mathit{modes}}$, and thus the optimized frequency band.

\begin{figure}[h!]
\centerline{\includegraphics[width=1\columnwidth]{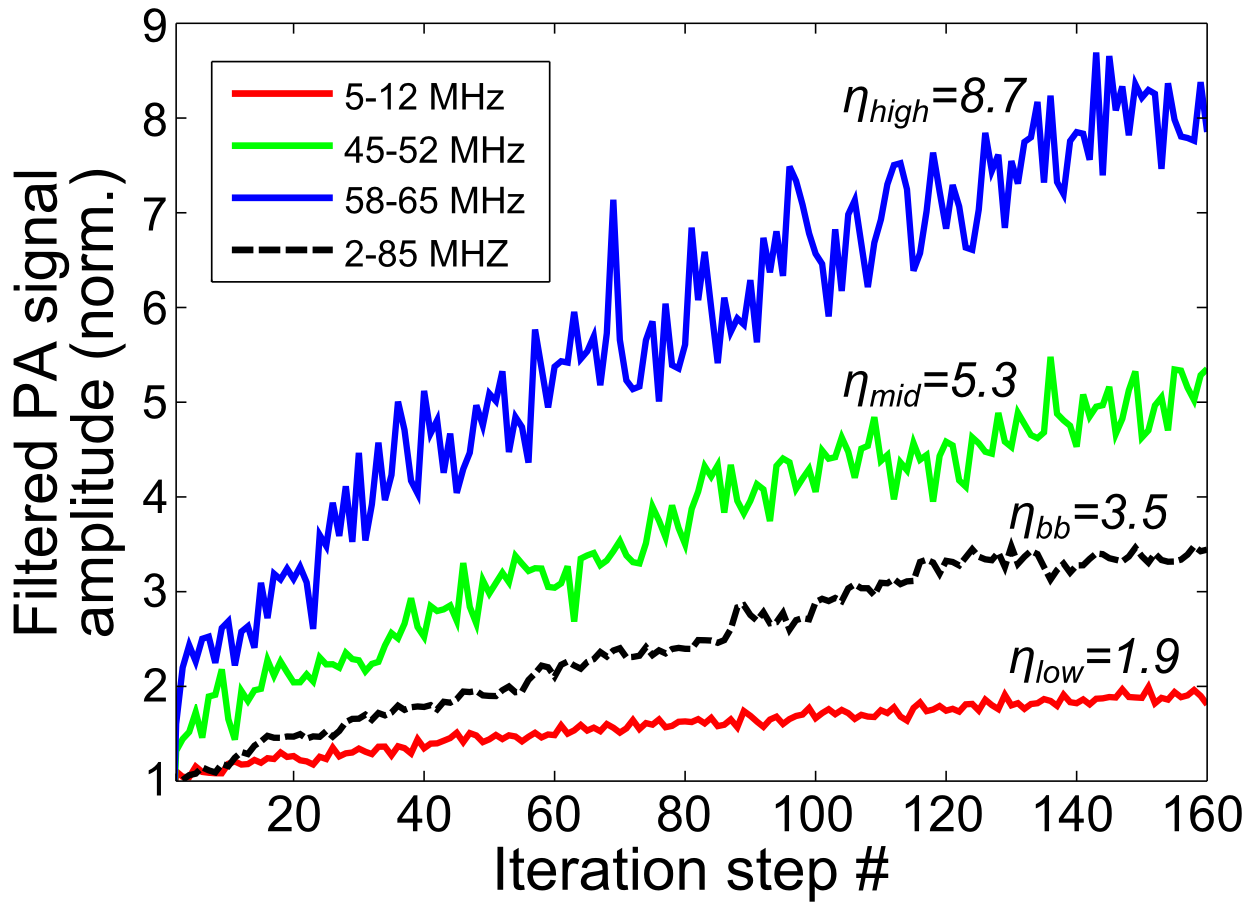}}
\caption{Evolution of the filtered photoacoustic feedback signals (RMS values) during their respective optimization (bb: broadband, 2-85MHz). The signals are normalized by the average of the signal over the first iteration of the optimization. 160 steps are needed, as this is the chosen Hadamard basis size, that spans the SLM pixels.}
\label{results}
\end{figure}

To experimentally demonstrate the influence of the spectral filtering, we optimized light delivery using different frequency bands. The calculated acoustic focal volumes corresponding to each band are illustrated in Fig.\ref{montage}c, illustrating the shrinking of the focus size with increasing detected frequency. The optimization results are presented in Fig.\ref{results}: the optimization of the high frequency band leads to a higher enhancement of the feedback signal than the optimization of lower frequency bands or the full band of the photoacoustic signal.
Specifically, the enhancement of the high frequency (48-65 MHz) photoacoustic signal is more than four times larger than the enhancement obtained during the optimization of the lower frequency content (5-12 MHz). 

It is important to note that shrinking the focal zone of the transducer results in a lower measured signal as the probed area is smaller and thus contains less speckle grains, which is expressed in a reduced signal-to-noise ratio in the high frequency trace in Fig.\ref{results}c (blue trace). In addition, limiting the signal bandwidth $\Delta f $ affects also the axial resolution of the acoustic probing $\Delta z = c_s/\Delta f $, and would have to be taken into account in the case of densely packed absorbing targets in the axial direction.  \newline

To relate our experimental observations to the spatial confinement of the optical focus, we have numerically modeled our system and run the same optimization algorithm. A cylindrical absorber of $30\mu m$ diameter and $1 mm$ length was simulated by considering photoacoustic point sources spaced by $2\mu m$ and excited by a 5 ns Gaussian laser pulse\cite{calasso2001photoacoustic}. The transducer geometry was modeled by discretizing the surface with $2.5\mu m$ steps. The values for the discretization steps were chosen small enough to provide an accurate modeling,  independent of the values themselves. The absorber was placed in the focal plane symmetrically to the transducer axis. The optical diffuser was modeled by a random phase mask (single scattering) so as to generate a speckle pattern with $25\mu m$ speckle grain size on the absorber, which is  placed in the diffuser's far field. The speckle pattern was assumed to be invariant in the propagation direction, i.e. along the width of the wire. The discrepency between the speed of sound in nylon and water ($c_{\mathit{nylon}}= 2620 m/s$) was taken into account by scaling the diamenter of both the wire and the speckle grain by a factor $\dfrac{c_{\mathit{water}}}{c_{\mathit{nylon}}}$.\\
\begin{figure}[h!]
\centerline{\includegraphics[width=1.1\columnwidth]{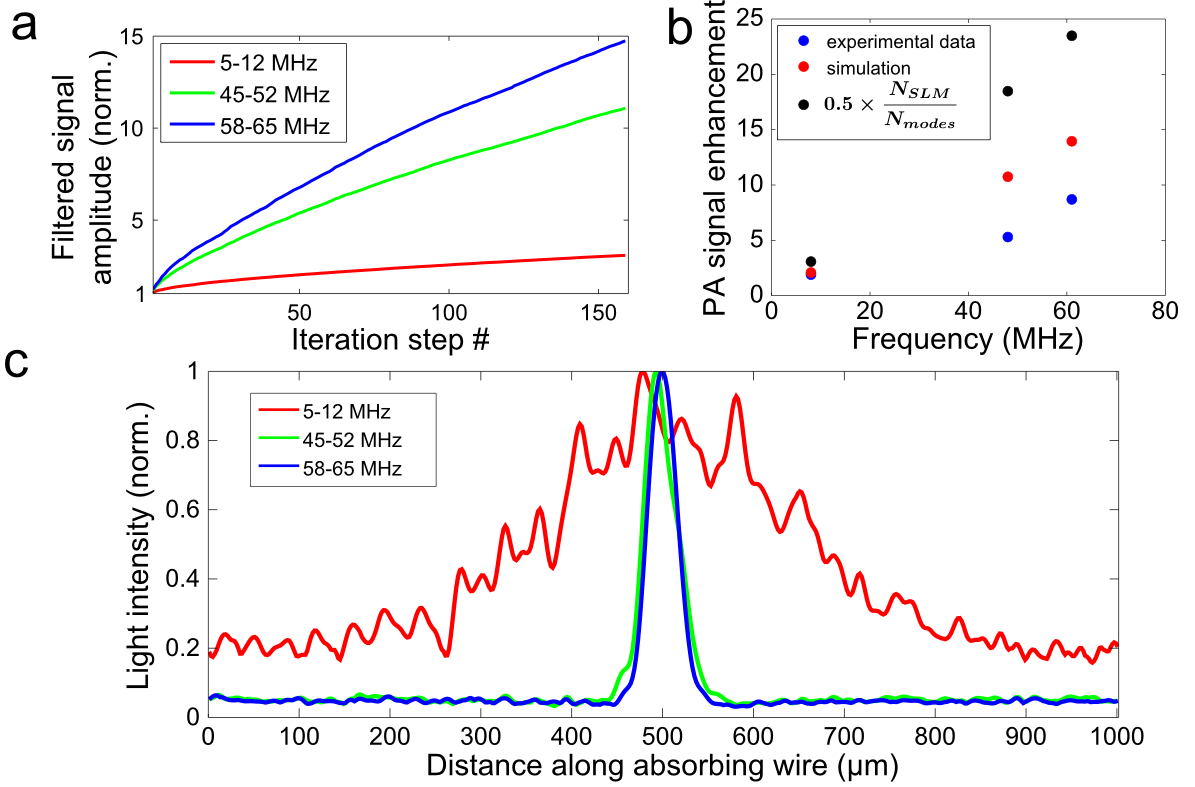}}
\caption{Simulation results: (a) Evolution of the PA signals during the optimization process. (b) Enhancement of spectrally filtered PA signal after optimization over different frequency bands. Experimental results follow the same trend than simulations. (c) Light intensity profile along the absorbing wire after optimization of the PA signal (averaged over several realisations of the initial speckle pattern).}
\label{fig4}
\end{figure}

Simulations results were averaged over 100 different realizations of the scattering medium. The average evolution of the PA feedback signals are presented in Fig.\ref{fig4}a. 

The light intensity profiles obtained after optimization were averaged over the initial speckle realizations and are presented in Fig.\ref{fig4}c. We observe that the optimization of the high frequency content of the photoacoustic signal leads to a tighter optical focus, as expected. 

Experiments and simulations follow the same trend (see Fig.\ref{fig4}b), but with a lower enhancement obtained experimentally. We attribute this fact to the lower experimental signal-to-noise ratio, especially for the high frequency bands, and to the experimental system stability..

We note that even the noise-free simulations do not exhibit the same behaviour than what would be obtained with a camera. This could be attributed to the fact that this approach does not take into account the geometry of the photoacoustic source. A smaller optical focus will indeed generate more high frequency than an extended one, which will affect the optimization of the filtered photoacoustic signal but not the optimization of the light intensity using a standard photodetector, even in the case where different weights would be attributed to different speckle grains.

In summary, we demonstrated that focusing light through scattering media by optimizing the photoacoustic response of an elongated cylindrical absorber can be improved optimizing the high frequency components of the photoacoustic signals.  
This type of elongated absorbing structures is an important model, relevant for mimicking blood vessels, a major endogenous source of photoacoustic constrast in tissue\cite{zhang2009vivo}.
The enhancement of the high frequency photoacoustic feedback has been experimentally and numerically shown to be larger than the optimization of the low frequency or content. We relate this improvement to the reduction in the acoustic focal volume of the spherically focused transducer as the frequency increases. 
We numerically linked these observations to the spatial confinement of light after optimization, showing that a tighter optical focus is indeed obtained after optimization of the high frequency components rather than the low frequency band of the photoacoustic signal. 
Spectrally filtered signals were chosen over peak-to-peak values (usually computed for photoacoustic imaging) because it enables precise control of the focal volume of the transducer. Indeed the frequency weighting in the peak-to-peak value is expected to change during the optimization, changing dynamically the size of the focal spot, which is prevented in our approach.

This new technique may become useful if one wants to achieve optical focusing inside scattering media, where the speckle grain size becomes indeed quickly small and diffraction limited. In such scenarios, one has to reduce the size of the acoustic focus as much as possible to get a measurable increase in the photoacoustic signal, using the smallest possible number of degrees of control (SLM pixels). 
The access to such spectrally controlled focus size does not have a direct counterpart in all-optical wavefront shaping experiments.\newline

\vspace{1cm}
\bibliography{optim}
\end{document}